\documentclass[conference]{IEEEtran}
\IEEEoverridecommandlockouts
\usepackage{cite}
\usepackage{amsmath,amssymb,amsfonts}
\DeclareMathOperator{\atan2}{atan2}
\usepackage{algorithmic}
\usepackage{graphicx}
\usepackage{textcomp}
\usepackage{xcolor}
\usepackage{booktabs}
\usepackage{listings}
\usepackage{amsthm}
\newtheorem{problem}{Problem}
\usepackage[justification=raggedright]{caption}

\usepackage[top=0.83in, bottom=0.85in, left=0.63in, right=0.63in]{geometry}

\def\BibTeX{{\rm B\kern-.05em{\sc i\kern-.025em b}\kern-.08em
    T\kern-.1667em\lower.7ex\hbox{E}\kern-.125emX}}
\begin{document}

\title{ Accelerated Digital Twin Learning for Edge AI: A Comparison of FPGA and Mobile GPU\\}

\author{\IEEEauthorblockN{Bin Xu, Ayan Banerjee, Midhat Urooj, Sandeep K.S. Gupta}
\IEEEauthorblockA{\textit{Impact Lab, School of Computing and Augmented Intelligence} \\
\textit{Arizona State University, Tempe, AZ, USA}\\
\{binxu4,abanerj3,murooj,sandeep.gupta\}@asu.edu}
}

\maketitle

\begin{abstract}
Digital twins (DTs) can enable precision healthcare by continually learning a mathematical representation of patient-specific dynamics. However, mission critical healthcare applications require fast, resource-efficient DT learning, which is often infeasible with existing model recovery (MR) techniques due to their reliance on iterative solvers and high compute/memory demands. In this paper, we present a general DT learning framework that is amenable to acceleration on reconfigurable hardware such as FPGAs, enabling substantial speedup and energy efficiency. We compare our FPGA-based implementation with a multi-processing implementation in mobile GPU, which is a popular choice for AI in edge devices. Further, we compare both edge AI implementations with cloud GPU baseline. Specifically, our FPGA implementation achieves an 8.8× improvement in \text{performance-per-watt} for the MR task, a 28.5×  reduction in DRAM footprint, and a 1.67×  runtime speedup compared to cloud GPU baselines. On the other hand, mobile GPU achieves 2x better performance per watts but has 2x increase in runtime and 10x more DRAM footprint than FPGA. We show the usage of this technique in DT guided synthetic data generation for Type 1 Diabetes and proactive coronary artery disease detection.
\end{abstract}


\section{Introduction}
\begin{figure*}[tbp]
\centering
\includegraphics[trim=0 0 0 0,width=\textwidth]{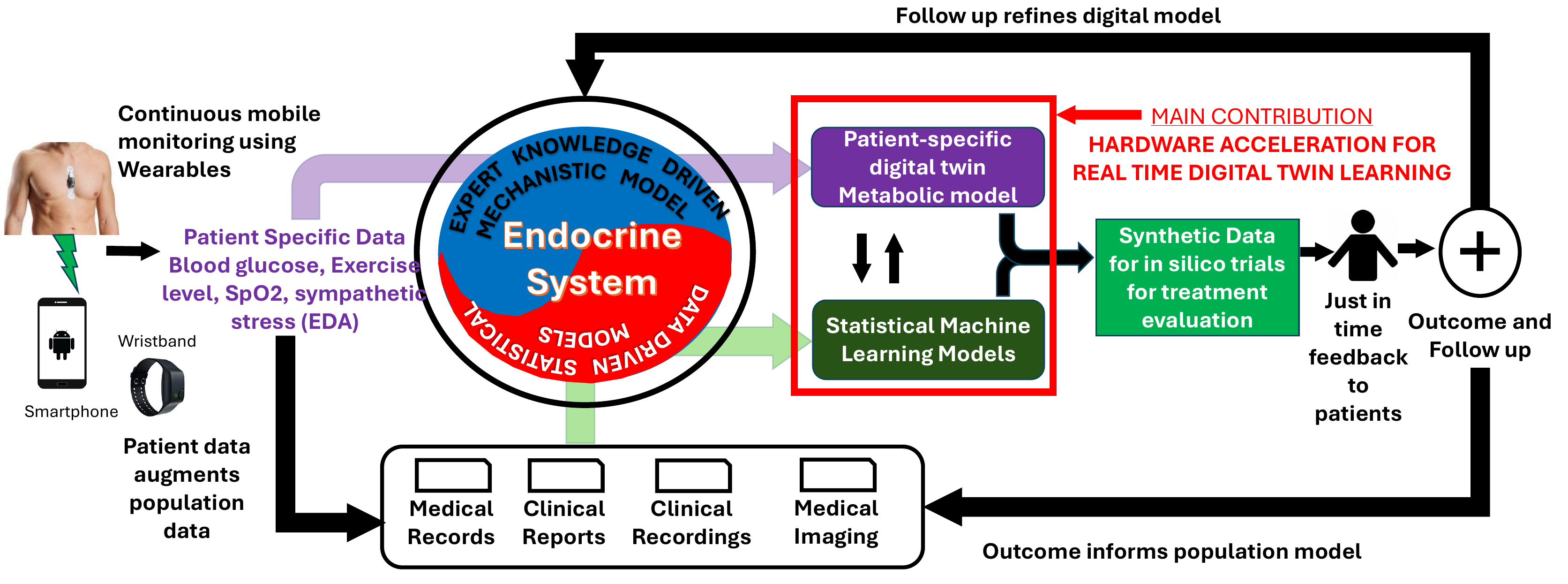}
\caption{Real-time digital twin learning for precision healthcare applications such as Diabetes Care.}
\label{fig:DTLearning}
\end{figure*}

A key technological innovation towards physical AI\cite{jensen2025keynote} is the concept of digital twin (DT). DTs are mathematical models of physical processes with two essential properties: a) the model structure is guided by first-principle satisfied by the physical process, and b) the model parameters are continuously calibrated with real world data in real-time. A major application of DT is in precision medicine, which brings a fundamental shift in disease management from decision making based on statistical inferences of individual variance of treatment efficacy to patient specific evaluations leading to just-in-time diagnosis, personalized treatment, and individualized recovery as shown in Figure \ref{fig:DTLearning}. 

A continuously calibrated DT can be used for simulating various potential treatment plans for their safety and efficacy on the specific patient~\cite{Banerjee12Ensuring,Banerjee12Percom,banerjee2015analysis,sadeghi2020system,LamraniTII}, derive a personalized verified safe and effective plan~\cite{Banerjee24ICPR,banerjee2024towards} or identify novel operational scenarios~\cite{Maity25TAI}. Given the mission critical nature of the application of precision medicine, the calibration, simulation and safety / efficacy feedback has to be performed within time constraints. These constraints are application specific and guided by hazard evolution dynamics~\cite{leveson2003applying} as shown in Table \ref{tbl:applications}. 

\subsection{Computational Challenges of DT Learning}
Automated continuous learning of DT in real-time is a major scientific challenge in the age of physical AI. 

\noindent{\bf Real-Time Challenge:} The primary computational component of DT learning is physics-guided model recovery~\cite{Banerjee2024}, where the model coefficients of a first-principle based differential dynamics is learned from real data under constraints of sampling, implicit or unmonitored dynamics, and human errors. The computational needs of physics-guided model recovery (MR) may prevent real-time operation even with parallelization with state-of-the-art (SOTA) multi-processing pipeline. Table \ref{tbl:applications} shows that the time to learn an application specific DT exceeds the response time required to avoid medical hazards. One of the fundamental reason is that analytical operations with physics-guided models require solution of differential dynamics which are iterative in nature. Such iterative operations are not amenable for parallelization. As such the SOTA multi-processing pipeline is less effective in real-time DT learning.    

\begin{table}[htbp]
\centering
\scriptsize
\caption{Use of digital twins to provide personalized feedback in precision healthcare applications. The table summarizes application-specific response time requirements and digital twin (DT) learning times on a SOTA GPU (NVIDIA RTX 6000).}
\setlength{\tabcolsep}{6pt}
\begin{tabular}{@{}p{0.5in}@{}p{0.05 in}@{}p{0.25 in}p{0.575 in}@{}p{0.3 in}p{0.7 in}@{}p{0.05 in}@{}p{0.3 in}@{}p{0.475 in}@{}}
\toprule
\textbf{Domain}& &\textbf{Hazard} & \textbf{Feedback} & \textbf{Response Time}&\textbf{DT model}& &\textbf{GPU Time} & \textbf{Data transfer time}\\
\midrule
Diabetes \cite{Banerjee2024}& &Hypo-glycemia& Change insulin rate &900s& Bergman minimal metabolic model&  
 &1412s& 27s\\
Cardiac disease~\cite{Salian24Asilomar,Nabar11GeMREM}& &Ischemia& Alert first responders &100s& ECG generative model& & 452s& 81s\\
Brain sensing~\cite{Banerjee2024,sadeghi2016toward}& &Attention deficit& Audio-visual cues&33ms& Resistance capacitance model& & 321s&125s\\
\bottomrule
\end{tabular}
\label{tbl:applications}
\end{table}

\noindent{\bf Edge AI\cite{singh2023edge} Challenge:} Data driven inferencing in real-time suffers from data transfer bottleneck (Table \ref{tbl:applications} shows data transfer times in medical DT applications forms a significant percentage or even exceeds response time). Recent advancements in edge AI aim to bring DT learning computation closer to data source, potentially bypassing the data transfer time. However, edge AI devices such as mobile GPU are resource constrained and hence may not be capable of DT learning within real-time constraints. 

\noindent{\bf Real World Challenge:} Data obtained from real world deployments of healthcare systems are restricted in sampling rate, and often compromised signal quality with potentially poor signal-to-noise ratio (SNR) especially when collected from human participants in free living conditions~\cite{pmlr-v255-banerjee24a,Banerjee24Sampling}. Moreover, privacy constraints may lead to unavailability of measurements of key dynamical parameters of the DTs. Hence, any DT learning mechanism in the real world require to learn implicit or unmeasured dynamics. Recently physics-guided sparse model recovery techniques such as Physics Informed Neural Networks (PINNs)~\cite{chen2021physics} or Physics informed Neural ODE (PiNODE)~\cite{PiNode} or Extracting sparse Model from ImpLicit dYnamics (EMILY)~\cite{pmlr-v255-banerjee24a} have been proposed to tackle implicit dynamics under low sampling frequencies. These techniques calibrate DT with real world data by following the Koopman theory~\cite{mauroy2020koopman}. The techniques attempt to learn a Koopman operator~\cite{mauroy2020koopman} that models the first-principle based DT dynamics using an expanded sparse state space where the dynamics become linear. The techniques utilize the universal function modeling capability of neural networks to learn the implicit dynamics while maintaining robustness to sensor noise. However, apart from the significant computational requirements of solving an Ordinary Differential Equation (ODE) in each learning step, these techniques also suffer from high memory requirements to store the expanded state space during computation. Hence, although these techniques are capable to calibrate DTs with real world data, they may not meet the real-time requirements and resource constraints of edge devices to support edge AI applications. 


\subsection{Contribution of manuscript}
The hardware acceleration of physics-guided model recovery techniques remains a relatively underexplored research area, particularly in terms of evaluating their feasibility for meeting the timing and resource constraints of edge AI applications.
In this paper, we demonstrate a pathway towards hardware acceleration of physics-guided model recovery such as PINNs, PiNODE, and EMILY to enable real-time DT calibration for precision healthcare. We show the application of hardware acceleration of DT learning for two exemplary medical applications on insulin management for Type 1 diabetes and coronary artery disease detection using electrocardiogram (ECG) sensors and show a feasibility analysis of hardware acceleration in meeting timing and resouce constraints of edge AI applications.

\section{Physics-guided Model Recovery}
\begin{figure}[tbp]
\centering
\includegraphics[width=\columnwidth,clip=false,trim=0 75 0 0]{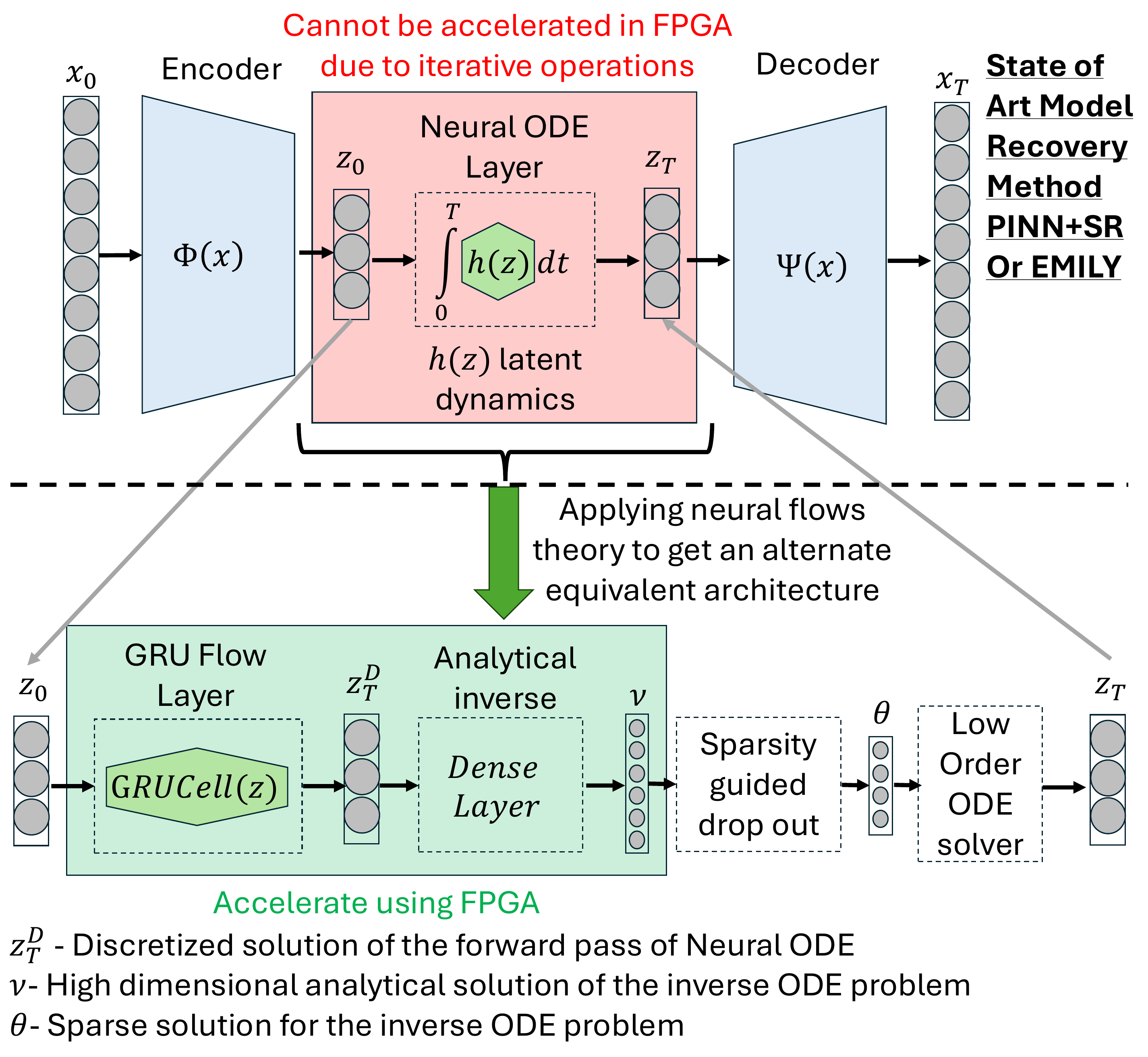}
\vspace{0.1in}
\captionsetup{justification=raggedright,singlelinecheck=false}
\caption{FPGA acceleration strategy\cite{bin2025modelrecovery} using neural flow-based equivalent architecture to neural ODEs.}
\label{fig:FPGA}
\end{figure}

The primary objective of MR is similar to an auto-encoder (Figure \ref{fig:FPGA}), where given a multivariate time series signal $X(t)$, the aim is to find a latent space representation that can be used to reconstruct an estimation $\Tilde{X}(t)$ with low error. It has the traditional encoder $\phi(t)$ and decoder $\Psi(t)$ of an auto-encoder architecture. MR represents the measurements $X$ of dimension $n$ and $N$ samples as a set of nonlinear ordinary differential equation model in Eqn: \ref{eqn:Model}. 
\begin{equation}
    \label{eqn:Model}
    \scriptsize
    \dot{X} = h(X,U,\theta),
\end{equation}
\vspace{-0.2in}\\
where $h$ is a parameterized nonlinear function, $U$ is the $m$ dimensional external input, and $\theta$ is the $p$ dimensional coefficient set of the nonlinear ODE model.

\noindent{\bf Sparsity:} An $n$-dimensional model with $M^{th}$ order nonlinearity can utilize $\binom{M+n}{n}$ nonlinear terms. A sparse model only includes a few nonlinear terms $p << \binom{M+n}{n}$. Sparsity structure of a model is the set of nonlinear terms used by it.


\noindent{\bf Identifiable model:} A model in Eqn.\  \ref{eqn:Model} is identifiable~\cite{verdiere2019systematic}, if $\exists$ time $t_I > 0$, such that $\forall \theta, \Tilde{\theta} \in \mathcal{R}^p$:
\begin{equation}
    \label{eqn:Ident}
    \scriptsize
    \forall t \in [0,t_I], f(X(t),U(t),\theta) = f(X(t),U(t),\Tilde{\theta}) \implies \theta = \Tilde{\theta}. 
\end{equation}

\noindent Eqn. \ref{eqn:Ident} effectively means that a model is identifiable if two different model coefficients do not result in identical measurements $X$. In simpler terms, this means $\forall \theta_i \in \theta, \frac{dX}{d\theta_i} \neq 0$. In this paper, we assume that the underlying model is identifiable.

\begin{problem}[Sparse Model Recovery]\label{prob:Problem} Given $N$ samples of measurements $X$ and inputs $U$, obtained from a sparse model in Eqn. \ref{eqn:Model} such that $\theta$ is identifiable, recover $\Tilde{\theta}$ such that for $\Tilde{X}$ generated from $f(X,U,\Tilde{\theta})$, we have $||X - \Tilde{X}|| \leq \epsilon$, where $\epsilon$ is the maximum tolerable error.
\end{problem}

\noindent{\bf Role of NODE:} Both EMILY~\cite{pmlr-v255-banerjee24a} and PINN~\cite{chen2021physics} utilize a layer of NODE cells in order to integrate the underlying nonlinear ODE dynamics. NODE cell's forward pass is by design the integration of the function $h$ over time horizon $T$ with $N$ samples (Fig. \ref{fig:FPGA}). This effectively requires an ODE solver in each cell of the NODE layer:
\vspace{-0.1in}\\
\begin{equation}
\scriptsize
z(t) = \int\limits^T_0{h(z,u,\theta)dt}, 
\end{equation}
where $z \in Z$ and $u \in U$ are each cells output and input. 
The results are then used further in the EMILY or PINN pipeline to extract the accurate underlying nonlinear ODE model.

\section{Hardware Acceleration of Physics-Guided MR}
A primary challenge in accelerating MR techniques lies in the iterative nature of ODE solvers, which are required for solving NODE cell operations during the forward pass. Recent works have explored the acceleration of standalone ODE solvers~\cite{stamoulias2017high,ebrahimi2017evaluation}, but these solutions assume fixed ODE model coefficients. Such fixed-coefficient approaches are not suitable for PiNODE, which requires solving a large number of ODEs with dynamically varying model coefficients, making traditional acceleration methods ineffective for generalizable MR architectures.
We leverage the theory of neural flows~\cite{bilovs2021neural} to develop an alternative neural structure that is mathematically equivalent to the NODE layers used in EMILY, PiNODE, and PINN while being more amenable to FPGA acceleration (Figure \ref{fig:FPGA}). Instead of using a conventional NODE layer, we apply a layer of invertible functions designed through a combination of Gated Recurrent Units (GRUs) and a dense layer of neurons with nonlinear activation functions. 
GRU~\cite{shiri2023comprehensive} are a type of recurrent neural network (RNN) architecture that introduces gating mechanisms to control the flow of information over time. Compared to traditional RNNs or LSTMs, GRUs are computationally efficient and require fewer parameters~\cite{chung2014empirical}, making them favorable for deployment on resource-constrained platforms such as FPGAs.

FPGA (Field-Programmable Gate Array) is a reconfigurable semiconductor device that enables developers to implement custom digital circuits directly in hardware~\cite{maxfield2004design}. Unlike fixed-function processors, FPGAs consist of an array of Configurable Logic Blocks (CLBs), Look-Up Tables (LUTs) for implementing combinational logic, Flip-Flops(FF) for sequential logic, and programmable interconnects~\cite{kuon2007measuring}. FPGAs also incorporate on-chip memory resources, such as Block RAM (BRAM) and UltraRAM (URAM), as well as Digital Signal Processing (DSP) slices optimized for arithmetic-intensive operations.

One of the primary challenges in FPGA design lies in efficiently mapping high-level algorithms onto limited hardware resources while maximizing performance. Loop-carried dependencies—such as Read-After-Write (RAW), Write-After-Read (WAR)—can inhibit effective pipelining, and reduce throughput. In addition to control hazards, memory access patterns pose a significant design challenge. FPGAs feature a hierarchical memory system including block RAM (BRAM), Look-Up Tables (LUTs), and registers(FF), all of which must be judiciously partitioned and scheduled to avoid access bottlenecks and ensure data locality.

In our design, we address these challenges through two key techniques: \textit{array partitioning} and \textit{loop pipelining}, both guided by high-level synthesis (HLS) directives. The FPGA kernel interfaces with the processor using an AXI4-Lite protocol, after which input data is transferred to on-chip memory. We apply full array partitioning using the directive \texttt{\#pragma HLS ARRAY\_PARTITION complete}, which instructs the HLS compiler to map each element of the input array to an independent storage resource—such as a dedicated register or BRAM segment. This partitioning strategy eliminates inter-element memory conflicts and enables parallel access to the data elements.

We then construct a fully parallelized model recovery pipeline on the FPGA. All major computational stages—including the forward pass, backpropagation, and loss computation—are pipelined using \texttt{\#pragma HLS PIPELINE II=1}. Once the inputs are partitioned and loop-carried dependencies are removed, this setup achieves an initiation interval (II) of 1, allowing a new iteration to begin every clock cycle. This significantly boosts throughput and latency performance. There can be the violation of loop dependency in the simulation. In order to eliminate RAW and WAR hazards, we need to test \texttt{\#pragma HLS PIPELINE II=2} or \texttt{\#pragma HLS PIPELINE II=3}, which means a new iteration begins in every 2 cycles and 3 cycles. If there is no time violation in the simulation, it means there are no RAW and WAR hazards. 
However, more cycles mean more latency in the pipeline of computation. 


\section{Edge AI Feasibility Evaluation}
We explore the feasibility of real-time DT calibration using real-world data in the edge for two exemplary applications of insulin management and electrocardiogram monitoring for coronary artery disease detection.

\subsection{Applications}\label{AA}
\noindent{\bf Automated insulin delivery (AID):} For the insulin management system the digital twin took the form of Eqn: \ref{eqn:5} - \ref{eqn:7} 

\begin{scriptsize}
\begin{eqnarray}
\label{eqn:5}
& &\dot{\delta i}(t) = -n\delta i(t) + p_4 u_1(t)\\
\label{eqn:6}
& & \dot{\delta} i_s(t) = -p_1 \delta i_s(t) + p_2 (\delta i(t) - i_b)\\
\label{eqn:7}
& & \dot{\delta} G(t) = -\delta i_s (t) G_b -p3 (\delta G(t)) + u2(t)/VoI,
\end{eqnarray}
\end{scriptsize}

The input vector $U(t)$ consists of the overnight basal insulin level $i_b$ and the glucose appearance rate in the body $u_2$. The output vector $Y(t)$ comprises the blood insulin level $i$, the interstitial insulin level $i_s$, and the blood glucose level $G$. In AP, only the blood glucose level $G$ is an measurable output. $i_s$ and $i$ are hidden states that are not measurable but contribute to the final glucose output. $p_1$, $p_2$, $p_3$, $p_4$, $n$, and $1/V_o I$ are all patient specific coefficients.

The DT was calibrated using the real-world OhioT1D dataset available in~\cite{marling2020ohiot1dm}. It is 14 time series data of glucose insulin dynamics. Each time series data had a duration of 16 hrs 40 mins which amounts to 200 samples of Continuous Glucose Monitor (CGM) and insulin data. 

\noindent{\bf Cardiac Digital Twin:}  The cardiac digital twin is based on the ECGSYN model\cite{gupta2013generative}, the state variables \((x,y,z)\) evolve as -

\begin{scriptsize}
\begin{eqnarray}
\dot{x} &= \bigl(1 - \sqrt{x^2 + y^2}\bigr)\,x - \omega(t)\,y, \\
\dot{y} &= \bigl(1 - \sqrt{x^2 + y^2}\bigr)\,y + \omega(t)\,x, \\
\dot{z} &= - \sum_{i\in\{P,Q,R,S,T\}}
            a_i\,\Delta\theta_i\,
            \exp\!\Bigl(-\tfrac{\Delta\theta_i^2}{2\,b_i^2}\Bigr)
          -\bigl(z - z_0(t)\bigr),
\end{eqnarray}
\end{scriptsize}where
$\Delta\theta_i = \bigl[\theta(t) - \theta_i\bigr]\bmod 2\pi,
\quad
\theta(t)=\operatorname{atan2}(y,x),
\quad
\omega(t)=\frac{2\pi}{r(t)},
\quad
z_0(t)=A_{\mathrm b}\sin(2\pi f_{\mathrm{resp}}t).$

Variables are defined as: [\(x,y\)] coordinates on the unit circle governing phase dynamics, [\(\theta(t)\)] instantaneous phase \(\atan2(y,x)\), [\(\omega(t)\)] angular velocity set by the instantaneous RR interval \(r(t)\), [\(z(t)\)] output signal, whose peaks form the ECG waveform, [\(\theta_i\)] angles of the P, Q, R, S, T peaks on the circle, [\(a_i\)] amplitude of the \(i\)th peak, [\(b_i\)] width (standard deviation) of the \(i\)th peak, [\(r(t)\)] time‐varying RR interval with prescribed power spectrum, [\(A_{\mathrm b},f_{\mathrm{resp}}\)]  baseline‐wander amplitude and respiratory frequency.

\subsection{Platforms Used}\label{AA}
To evaluate the performance of our FPGA-based accelerator, we conducted a series of experiments across three hardware platforms: a cloud GPU, an edge-based mobile GPU, and a resource-constrained FPGA. The cloud GPU serves as the baseline for comparison, with a focus on evaluating power efficiency, execution time, and inference accuracy across all platforms.

\noindent\textbf{GPU Platform:}  
Experiments were first conducted on a workstation equipped with an Intel Xeon w9-3475X CPU and an NVIDIA RTX 6000 GPU with 48~GB of memory. Models were implemented using TensorFlow 2.10 and Keras 2.10. Power consumption was monitored using \texttt{nvidia-smi}, while execution time and DRAM footprint were recorded using the \texttt{time} and \texttt{psutil} libraries.

\noindent\textbf{Mobile GPU Platform:}  
To assess edge-level performance, we deployed models on the NVIDIA Jetson Orin Nano Developer Kit. This platform features a 6-core Arm Cortex-A78AE CPU and 8~GB of LPDDR5 memory. Its integrated GPU, based on the NVIDIA Ampere architecture, includes 1024 CUDA cores and 32 Tensor Cores. Power consumption was measured using \texttt{tegrastats}.

\noindent\textbf{FPGA Platform:}  
For FPGA implementation, experiments were performed on the PYNQ-Z2 board, which includes a dual-core ARM Cortex-A9 processor and a 1.3M-configurable-gate FPGA. The GRU model was developed from scratch, with both forward pass and backpropagation logic implemented in C++ using High-Level Synthesis (HLS) in AMD’s Vitis toolchain. The forward-pass accelerator was integrated using Direct Memory Access (DMA) to interface with the processing system. Power consumption was evaluated through Vivado's power analysis, while runtime and DRAM usage were recorded using the \texttt{time} and \texttt{psutil} libraries. All \texttt{\#pragma} directives and hardware-specific constructs used are compliant with Vitis High-Level Synthesis (Vitis HLS)~\cite{amdamd}, and the design was compiled using the Vitis HLS compiler.

\subsection{Results}
As shown in Table~\ref{tbl:mr_mode_performance} and Table \ref{tbl:mr_mode_performance_ecg}, the FPGA implementation achieves substantial efficiency gains compared to the GPU baseline. Specifically, it offers an 8.8× improvement in performance-per-watt for the MR task and achieves over 28.5× reduction in DRAM footprint. Additionally, the FPGA provides a 1.67× speedup in runtime for MR, despite operating at significantly lower frequencies. These results align with the comprehensive work by Cong et al. \cite{cong2018understanding}, which compare the performance of FPGAs and GPUs across a variety of application domains.

\begin{figure}[thbp]
\centering
\includegraphics[width=\columnwidth]{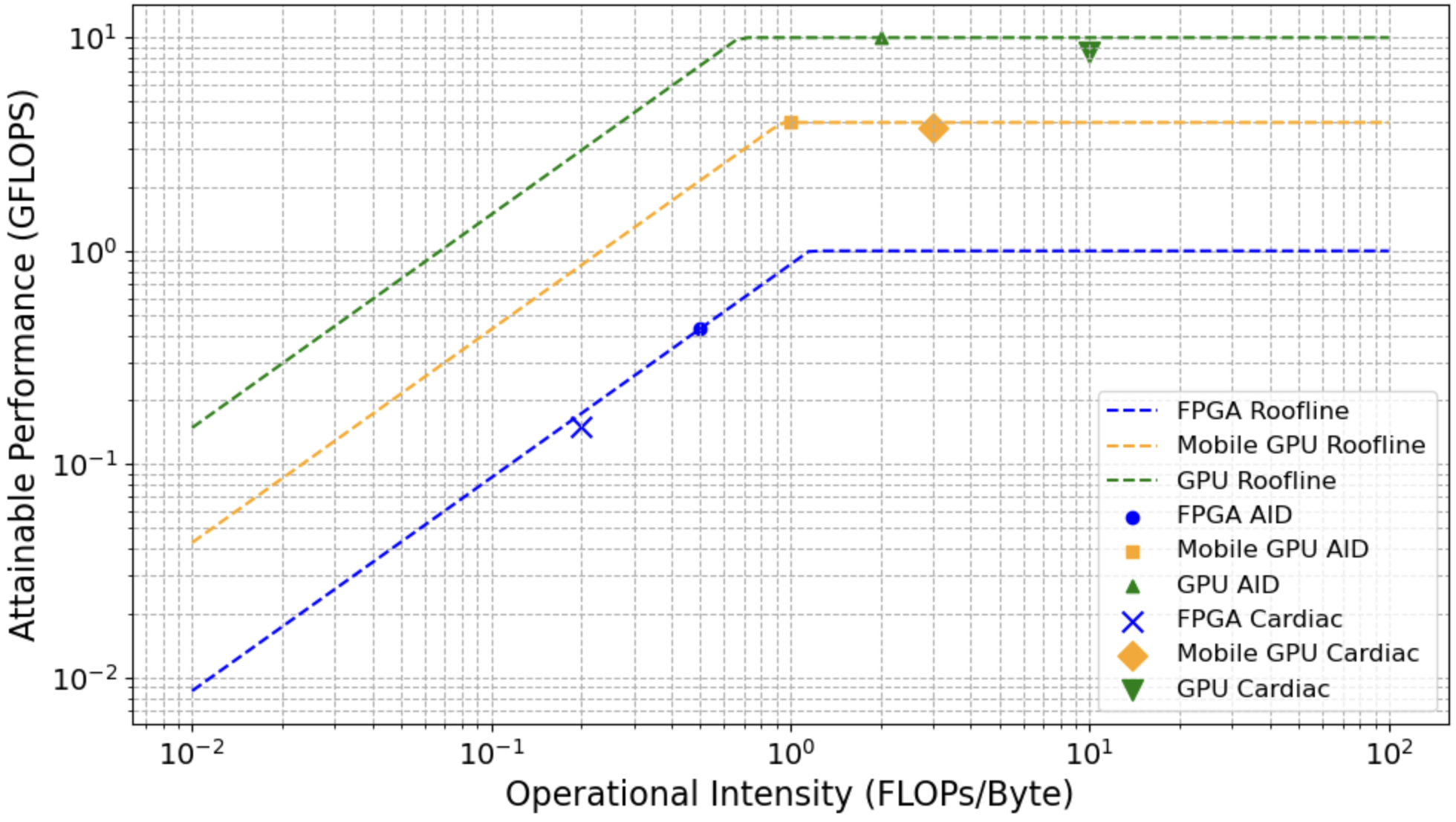}
\captionsetup{justification=raggedright,singlelinecheck=false}
\caption{Roofline model across FPGA, mobile GPU, and cloud GPU platforms, plotted on log-log scale.}
\label{fig:roofline}
\end{figure}

Figure~\ref{fig:roofline} illustrates the roofline model\cite{ofenbeck2014applying} comparison among FPGA, mobile GPU, and cloud GPU platforms for the MR task. The FPGA demonstrates efficiency at low operational intensities (around 0.5 FLOPs/Byte), operating near its memory bandwidth limit. This aligns well with the demands of real-time, low-batch-size workloads typical in model recovery applications.
Although the FPGA has a lower compute ceiling (1 GFLOPS) compared to the cloud GPU (10 GFLOPS), it delivers over 8× better performance-per-watt and achieves 1.6× faster runtime under the same task. Cloud GPUs, while powerful, are optimized for high operational intensities and suffer inefficiencies when handling memory-bound edge workloads. Mobile GPUs strike a balance between the two but still require significantly higher DRAM footprint than FPGAs to reach similar runtime performance.

From the Figure~\ref{fig:roofline}, we find that applications that benefit most from FPGA implementation under these constraints are those that are memory-bound, latency-sensitive, and have modest compute requirements. Examples include real-time physiological signal processing, streaming sensor fusion, and lightweight edge inference where throughput per watt and deterministic latency are more critical than peak FLOPs.

\begin{figure}[thbp]
    \centering
    \includegraphics[width=0.95\columnwidth]{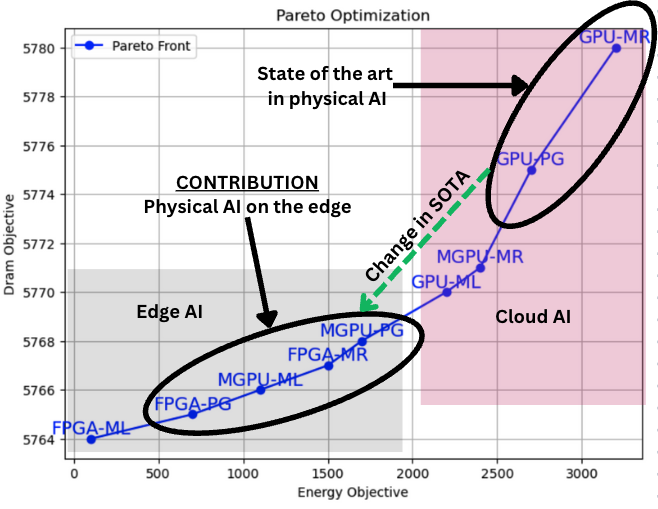}
    \captionsetup{justification=raggedright,singlelinecheck=false}
    \caption{Each blue dot represents a Pareto-optimal solution spanning Machine Learning (ML), Physics-Guided Machine Learning (PG), and Model Recovery (MR) tasks across FPGA, Mobile GPU (MGPU), and GPU platforms. State-of-the-art physical AI typically relies on cloud GPUs, represented in the red region. This work enables physical AI on edge platforms—illustrated in the grey region—achieving substantial reductions in both DRAM footprint and energy consumption.}
    \label{fig:pareto_optimization}
\end{figure}

We derived the Pareto front from experimental results corresponding to the optimal hyperparameter settings of the hardware acceleration strategy. Figure~\ref{fig:pareto_optimization} illustrates the Pareto front spanning Machine Learning (ML), Physics-Guided Machine Learning (PG)~\cite{pawar2021physics}, which integrates physical laws or domain knowledge into data-driven models, and Model Recovery (MR) tasks across FPGA, Mobile GPU (MGPU), and GPU platforms.
The trend shows a clear separation between edge AI and cloud AI and shows the feasibility of an FPGA-based solution to achieve high speed and lower energy consumption with a modest DRAM footprint, making them ideal for DT learning in the edge. In contrast, GPU-based solutions require higher power but offer greater memory bandwidth, making them more suitable for compute-intensive workloads. The performance of MGPU falls between that of the FPGA and cloud GPU.



\begin{table}[thbp]
\centering
\scriptsize
\caption{Performance comparison among FPGA, Mobile GPU, and GPU for \textbf{AID}.}
\begin{tabular}{lccc}
\toprule
\textbf{Metric} & \textbf{FPGA} & \textbf{Mobile GPU} & \textbf{GPU} \\
\midrule
\textbf{Average Error} & 4.60 & 3.07 & \textbf{2.90} \\
\textbf{Runtime (s)} & \textbf{253.84} & 562.75 & 423.28 \\
\textbf{Avg Power (W)} & \textbf{4.905} & 5.532 & 72.00 \\
\textbf{DRAM Footprint (MB)} & \textbf{214.23} & 2355.13 & 6118.36 \\
\textbf{Freq (MHz)} & 173 & 306 & \textbf{1410} \\
\textbf{Perf/Watt (s/W)} & 51.76 & \textbf{101.74} & 5.88 \\
\bottomrule
\end{tabular}
\label{tbl:mr_mode_performance}
\end{table}



\begin{table}[thbp]
\centering
\scriptsize
\caption{Performance comparison among FPGA, Mobile GPU, and GPU for \textbf{Cardiac}.}
\begin{tabular}{lccc}
\toprule
\textbf{Metric} & \textbf{FPGA} & \textbf{Mobile GPU} & \textbf{GPU} \\
\midrule
\textbf{Average Error} & 7.20 & 8.07 & \textbf{6.10} \\
\textbf{Runtime (s)} & \textbf{25.6} & 115.2 & 103.4 \\
\textbf{DRAM Footprint (MB)} & \textbf{121.3} & 1061.3 & 3101.6 \\
\textbf{Perf/Watt (s/W)} & 5.22 & \textbf{20.83} & 1.44 \\
\bottomrule
\end{tabular}
\label{tbl:mr_mode_performance_ecg}
\end{table}


\section{Discussion and Future Directions}

At its core, our work embodies the philosophy that algorithm and hardware must co‐evolve: by reformulating physics‐guided ODE solvers into neural‐flow–inspired GRU+Dense blocks, we unlock a class of learning models whose data dependencies map naturally onto deep pipelines and fine‐grained parallelism.  This co‐design approach transforms inherently iterative recovery methods into streamable computations, closing the gap between mathematical expressivity and real‐time edge deployment.  Beyond raw performance metrics, it reframes digital twins as living, on-device entities—no longer tethered to cloud resources but capable of continuous, patient-specific adaptation in situ.

Looking forward, this hardware‐acceleration paradigm will extend in two key directions.  First, heterogeneous integration of mixed-signal and analog compute elements promises even greater energy efficiency for sparse dynamics, enabling miniaturized accelerators in wearables and implantables.  Second, automated hardware synthesis from high-level model descriptions—leveraging domain-specific languages and compiler frameworks—will democratize custom accelerator design, so that new physiological models can be instantly mapped to optimized circuits.  Together, these advances will drive a new generation of truly autonomous, safety-critical digital twins that learn, infer, and adapt at the edge.

\section*{Acknowledgments}
The work is partially funded by NSF FDT-Biotech grant 2436801.

\bibliographystyle{ieeetr}
\bibliography{ref}

\end{document}